\documentstyle[psfig,floats,twocolumn,aps,prl]{revtex}
\begin{document}
\ifx\undefined\psfig\def\psfig#1{ }\else\fi
\ifpreprintsty\else
\twocolumn[\hsize\textwidth%
\columnwidth\hsize\csname@twocolumnfalse\endcsname
\fi
\draft
\preprint{ }
\title {Bosonization of the two-dimensional electron gas in the lowest
  Landau level} 
\author  {S. Conti}  
\address {Max-Planck-Institute for Mathematics in the Sciences,
  04103 Leipzig, Germany}
\author  {G. Vignale} 
\address{Department of Physics,  University of Missouri, Columbia,
  Missouri 65211}
\date{\today} 
\maketitle

\begin{abstract}
We develop a bosonization scheme for the collective  dynamics of a 
spinless two-dimensional electron gas  (2DEG) in the lowest Landau level.
The system is treated as a continuous elastic medium, and quantum
commutation relations are imposed between orthogonal components of the
elastic displacement field.  This theory provides a unified description
of bulk and edge excitations of compressible and incompressible phases,
and explains the results of recent tunneling experiments at the edge
of the 2DEG.
\end{abstract} 
\ifpreprintsty\else\vskip1pc]\fi
\narrowtext
Bosonization \cite{Bosonization} is a powerful mathematical technique to
describe the long-wavelength dynamics of correlated electronic systems,
in which the  motion of the electrons is highly collectivized. 
This technique has proved very useful in studies of the behavior of
interacting electrons in one-dimensional metals. 
Higher-dimensional generalizations have also been proposed
\cite{FradkinLuther}.  More recently, it has been realized that the 
{\it edges} of a two-dimensional electron gas (2DEG) in a strong
magnetic field  behave as a ``chiral Luttinger liquid"\cite{Wen}  and are
therefore amenable to bosonization \cite{Wen,Orgad}.  A collective
description of the {\it uniform} incompressible 2DEG in a strong
magnetic field such that all the electrons reside within the lowest
Landau level (LLL) \cite{Laughlin}  was pioneered by Girvin {\em et al.}
\cite {Girvin86}, and an effective bosonic theory was developed shortly
afterwards \cite{ZhangRead}. The similarity between the local
structure of the 2DEG in the LLL and that of a classical Wigner crystal
was exploited by Johansson and Kinaret \cite{JohanssonKinaret94} 
to set up an ``independent boson model"  for the electronic spectral
function; an equivalent model was derived by Haussmann
\cite{Haussmann96} from diagrammatic theory. Finally, several
authors have used bosonization schemes to describe the
dynamics of a smooth compressible edge in a quantum Hall
liquid \cite{Aleiner94,CV96CV97drs,HanThouless}, or of the electron gas
in a weak magnetic field \cite{Aleiner95}.

In this Letter we formulate a bosonic
theory of the linear dynamics of a {\it general} uniform or nonuniform
distribution of electrons in the LLL. 
Unlike previous theories
\cite{ZhangRead,JohanssonKinaret94,Haussmann96,Aleiner94,CV96CV97drs,HanThouless,Aleiner95},
the present one treats 
bulk and edge excitations of compressible and incompressible
states on an equal footing. 
One of our main results is a simple explanation for
recent experimental observations of one-electron tunneling characteristics
in the edge \cite {Chang} of the 2DEG.

The basic idea of our approach is to
treat the electrons in the LLL as a continuous elastic medium
characterized by an 
equilibrium density $\rho_0(\vec r)$, and by a local displacement
field $\vec u(\vec r,t)$, such that the time-dependent density is (to
first order in $\vec u$) $\rho (\vec r,t) = \rho_0(\vec r) - \vec \nabla
\cdot[\rho_0(\vec r) \vec u (\vec r,t)] \equiv \rho_0(\vec r) + \delta
\rho(\vec r,t)$. We ignore the spin in this paper. 
Under the assumption that both the equilibrium density and the displacement are
{\it   slowly varying} on the scale of the magnetic length $l = (\hbar c/e
B)^{1/2}$ (where $B$ is the magnetic field) the effective long-wavelength
Hamiltonian is 
\begin {eqnarray}
  H \!&=& {1 \over 2}\! \int \!\!d\vec r\! \int \!\!d \vec {r'} \delta
\rho(\vec r)
\left ( {e^2 \over \vert \vec r -
    \vec {r'} \vert} - \tilde \chi^{-1}(\vec r, \vec {r'}) \right ) \delta
\rho(\vec{r'}) \nonumber \\
  && +
  \int d\vec r \mu\left[\rho_0(\vec r)\right]
 \sum_{\alpha,\beta}\left[ s_{\alpha\beta}(\vec r) - {1\over2}
  \delta_{\alpha\beta} \vec\nabla\cdot\vec u(\vec r)\right]^2\,,
\label {Hamiltonian}
\end{eqnarray} 
where $\tilde \chi^{-1}(\vec r,\vec r')$ is the inverse of the {\em proper}
static density-density response function \cite{Nozieres}, $\mu$ is the
shear modulus (discussed below),  $s_{\alpha\beta} \equiv [\partial
u_\alpha (\vec r) 
/ \partial r_\beta + \partial u_\beta (\vec r) / \partial r_\alpha]/2$
is the strain tensor, and $\alpha,\beta$ are cartesian
indices. Eq.~(\ref{Hamiltonian}) generalizes the Hamiltonian of
classical elasticity theory 
\cite{Landau7}  to take into account the
long range of the Coulomb interaction, and of $\tilde\chi (\vec r,
\vec {r'})$ in the incompressible states  of the 2DEG \cite {Girvin86}.

The second term of Eq.~(\ref{Hamiltonian}) is the crucial new feature of
our theory. Ordinarily, the shear modulus of a liquid is taken to be
zero, but this is only true for shear deformations which occur slowly
compared to the  thermalization time $\tau$ of the
system. At frequencies higher than $1/\tau$ 
the correlated electron liquid behaves as a  Wigner glass, and
therefore exhibits a nonvanishing shear modulus. The essential
assumption is that, 
in this system, the relaxation frequency scale $1/\tau$ is much
smaller  than the characteristic frequency of the collective
dynamics.

The algebra of the displacement operators $u_x$, $u_y$, in the LLL,
is deduced from the canonical quantization condition
for the hydrodynamical momentum and displacement fields $[p_\alpha(\vec
r), u_\beta(\vec {r'})]= -i \delta (\vec r - \vec {r'}) \delta_{\alpha
\beta}$, by projecting out the higher Landau levels, i.e. averaging
over the fast cyclotron motion. This leads to
\begin {equation} 
  [u_\alpha(\vec r),u_\beta(\vec{r'})] = -i
  \epsilon_{\alpha\beta} \delta(\vec r - \vec{r'}) \, l^2  /
  \rho_0(\vec r)\,, 
\label{commutator} 
\end {equation}
where $\epsilon_{\alpha\beta}$ is the two-dimensional Levi-Civita
tensor. 
 This is consistent with the commutation relation between
projected density fluctuations \cite{Girvin86} {\it in the long
  wavelength limit}, namely $[\rho (\vec r),\rho(\vec {r'})] = - i \vec
\nabla_r \times \vec \nabla_{r'}\rho_0(\vec r) l^2 \delta(\vec r -
\vec {r'})$.

Bosonization of (\ref{Hamiltonian}) is accomplished by 
\begin {equation}
  u_\alpha(\vec r) = {l \over \rho_0^{1/2}(\vec r)}\sum_{n>0}   
  \left[ b_n g^*_{n\alpha}(\vec r) + b_n^{\dagger} g_{n\alpha}(\vec
    r)\right]\,,
\label {bosonization}
\end {equation} 
where $b_n$ and
$b_n^{\dagger}$ are boson 
operators, satisfying the usual algebra
$[b_n^{},b_{n'}^\dagger]=\delta_{nn'}$ etc..  The functions
$g_{n\alpha}(\vec r)$ ($g^*_{n \alpha}(\vec r)$), with $n>0$, are the
positive (negative) frequency solutions of the eigenvalue problem
\begin {equation} 
  \int H_{\alpha\beta}(\vec r, \vec{r'})
g_{n\beta}(\vec{r'}) d \vec{r'} =  \omega_n g_{n\alpha}(\vec r)
\label {eigenvalueproblem} \end{equation}
which is equivalent to the classical equation of motion.
(We adopt the convention of summing over repeated cartesian indices.) The
operator $H_{\alpha \beta}$ is given by 
\begin {eqnarray} H_{\alpha \beta}(\vec
  r, \vec{r'}) &=& -i l^2\rho_0^{1/2}(\vec r) 
  \epsilon_{\alpha\gamma}\left[\partial_\gamma
    \partial'_\beta \chi^{-1}(\vec r,\vec {r'})\right]
  \rho_0^{1/2}(\vec {r'})
  \nonumber\\
  &&\hskip-20mm +i{l^2\over\rho_0(\vec r)}
\epsilon_{\alpha\gamma} \left[ \partial^{}_\beta
    \partial'_\gamma
-\partial^{}_\gamma \partial'_\beta + \delta_{\gamma \beta}
    \partial^{}_\eta \partial'_\eta\right] \left[ \delta(\vec r -
    \vec {r'}) \mu \right],  \label{H} \end {eqnarray} 
where $\chi^{-1}(\vec r,\vec {r'}) \equiv \tilde \chi^{-1}(\vec r,\vec
{r'}) -e^2/\vert \vec r - \vec {r'} \vert$, and $\partial_\alpha$ and
$\partial'_\beta$ denote derivatives with respect to $r_\alpha$ and
$r_\beta'$ respectively.  It is hermitian with respect to the scalar
product   $(f,g) \equiv \int f^*_\alpha (\vec r) i \epsilon_{\alpha
\beta}   g_\beta(\vec r)d \vec r$,
and  its eigenvalues are therefore real. For each eigenfunction $g_n$
with positive eigenvalue $\omega_n$, there is a complex conjugate
one $g_n^* \equiv g_{-n}$ with negative eigenvalue $-
\omega_n$. These eigenfunctions can be chosen to satisfy orthonormality
relations $(g_n,g_m) = sgn(n) \delta_{nm}$ ($n$ and $m$ can now have
either sign) \cite{Footnote11}. The ``completeness relation'' has the
form \begin {equation}
  \sum_{n>0}[ g^{}_{n\alpha}(\vec r) g^*_{n\beta}(\vec{r'})-
  g^*_{n\alpha}(\vec r) g^{}_{n\beta}(\vec{r'})] = i \epsilon_{\alpha
    \beta} \delta(\vec r - \vec{r'}).
\end {equation}
These equations guarantee that the commutation relation
(\ref{commutator}) is satisfied.  Then, substitution of
Eq.~(\ref{bosonization}) into Eq.~(\ref{Hamiltonian}) yields the
Hamiltonian in the desired form $ H = \sum_{n>0}\left(b^{\dagger}_n
b_n^{} + 1/2 \right)   \omega_n$.
    
Our next step is to construct
the electron  tunneling operator $\psi(\vec R$) associated with the LLL
coherent state orbital centered 
at $\vec R$.   This operator must satisfy  the
commutation relation
\begin {equation}
  [\psi(\vec R), \rho(\vec r)] = \psi (\vec R) \exp[-\vert \vec
r-\vec R\vert^2/2 l^2]/2\pi l^2\,, 
\label{eqcompsirho}
\end {equation} 
i.e., destroy a  gaussian density near $\vec R$.
Eq.~(\ref{eqcompsirho}) alone does not uniquely determine 
$\psi(\vec R)$. In order to do this we must also specify its
commutator with the vorticity $\rho_v(\vec r) \equiv - \epsilon_{\alpha
\beta} \partial_\alpha[ \rho_0(\vec r) u_\beta(\vec r)]$. 
For the purpose of studying tunneling, we choose
$[\psi(\vec R),\rho_v(\vec r)]=0$, because 
the incoming electron is not expected to change the vorticity at the
initial time.  The solution has the form
\begin {equation}
  \psi(\vec R) = \exp\left[-\sum_{n>0}\left({M^*_n(\vec R) \over
\omega_n}b_n
      -{M_n(\vec R) \over \omega_n}b_n^{\dagger}\right)\right]
\label {psi} \end {equation} 
where the ``electron-phonon" matrix elements $M_n(\vec R)$ are written
as $M_n(\vec R)= \int\exp[-\vert \vec
r-\vec R\vert^2/2 l^2]\tilde M_n(\vec{r}) d\vec
r/2\pi l^2$, and  
\begin {equation}
{\tilde M_n(\vec r) \over \omega_n} = {i \over 2 \pi} \int
\epsilon_{\alpha \beta}  {g_{n\alpha}(\vec {r'}) \over \rho_0^{1/2}(\vec
  {r'}) l}
\partial_\beta \log \vert \vec r - \vec {r'} \vert
d \vec {r'}  \,.
\label {Mtilde}
\end {equation}

The calculation of the local Green's function 
$G(\vec R,t) = - i \langle T \psi(\vec R,t) \psi^{\dagger}(\vec R,0)
\rangle$ can be carried out by standard techniques 
\cite {Mahan,JohanssonKinaret94}.  Here
we simply report the zero temperature result for the integral equation
connecting the electronic spectral function $A(\vec R,\omega)$ (the
Fourier transform of $G(\vec R,t)$) to the collective
excitation spectrum:
\begin {equation} \omega A(\vec R,\omega) = \int_0^\omega 
  g(\vec R,\Omega)\,A(\vec R,\omega-\Omega) 
d \Omega
\label{A}\end {equation} where
$\omega >0$ and
\begin {equation}
  g(\vec R,\Omega) = {1 \over \Omega}\sum_{n>0} \vert M_n(\vec R) \vert ^2
  \delta(\Omega- \omega_n) \label{g}
\end {equation}
is closely related to the local dynamical structure factor of the
liquid.  The above equations constitute a complete scheme for the
calculation of single particle and collective properties of a general
distribution of electrons in the LLL.
We now discuss some specific examples.

(1) {\it The uniform 2DEG}.  Let $\rho_0$ be the uniform density. The 
normalized eigenfunctions of Eq.~(\ref{eigenvalueproblem}) have the
form 
\begin {equation}
  g_{\vec q L}(\vec r) = i ql\left({\mu \over 2 \omega_q \rho_0}
  \right )^{1/2} e^{i \vec q \cdot \vec r}, \label {gL} \end{equation}
and
\begin {equation} g_{\vec q T}(\vec r) ={1 \over ql} \left
    ({\omega_q \rho_0 \over 2 \mu} \right )^{1/2} e^{i \vec q
    \cdot \vec r}, \label {gT} 
\end{equation} 
where
\begin {equation} \omega_q = \mu^{1/2}
\left (v(q) +{K(q)+\mu \over \rho_0^2} \right )^{1/2} (ql)^2 
\label {dispersion} 
\end {equation}
is the frequency of the mode at wavevector $\vec q$, $v(q)$ is the
Fourier transform of the electron-electron interaction, $K(q) = -
\rho_0^2 \tilde \chi^{-1}(q)$ is the $q$-dependent bulk modulus, and the
labels $L$ (longitudinal) and $T$ (transverse) refer to the components
parallel and perpendicular to $\vec q$.
The electron-phonon matrix element of Eq.~(\ref{Mtilde}) has the form
\begin {equation}
  M_{q}(\vec r) = \left[v(q) + {K_{\vec r}(q) + \mu \over
     \rho_0^2}\right]  e^{-q^2 l^2/2} 
\rho_q(\vec r) \,,
\label{Mtildehomogeneous}  \end {equation} 
where $\rho_q \equiv -iql
\rho_0^{1/2}g_{qL}(\vec r)$ and $K_{\vec r}(q) \equiv -\rho_0^{2} \int
\tilde \chi^{-1}(\vec r - \vec {r'})\exp[i \vec q \cdot (\vec {r'}-\vec
r)]d\vec {r'}$. (The reason for the apparently unnecessary subscript
$\vec r$ will become clear below). 
 
\begin{figure}
\centerline{\psfig{figure=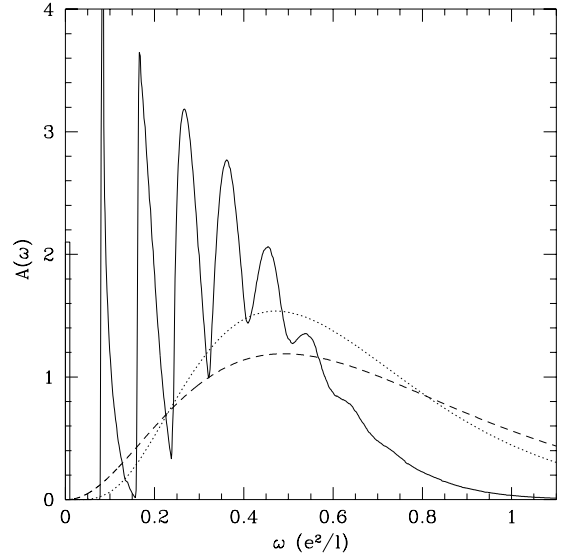,width=0.9\columnwidth}}
\caption{Spectral function for a compressible liquid at $\nu=0.3$ with
  short range interactions ($d=10 l$, dashed curve), with long-range
  interactions (dotted curve) and for an incompressible liquid at
  $\nu=1/3$ (full curve, the $\delta$-function peak at $\omega=0$
  contains around $2\%$ of the spectral strength).} 
\label{fig1}
\end{figure}

We now distinguish two cases: {\it (i)  Compressible case.}
$\lim_{q \to 0} K(q) = K$ is finite.  The values of $K$ and $\mu$ can be
approximated by those of a classical Wigner
crystal, namely, $\mu \sim \mu_c = 0.09775 \rho_0
(e^2/l) \nu^{1/2}$  and $K \sim K_c = 6 \mu_c$ \cite{Maradudin}. 
The long wavelength modes have frequencies $\omega_q \propto
q^{3/2}$ for Coulomb-like interaction ($v(q)
= 2 \pi e^2/q$) and $\omega_q \propto q^2$ for
short-range interaction ($v(q) = e^2d$). We have now all the necessary
input for calculating the spectral function from
Eqs.~(\ref{A},\ref{g}). The 
low-frequency behavior (where ``zero" frequency is the chemical
potential) is found to be $A(\omega) \propto \omega^{-5/4} e^{-
\gamma/\omega^{1/2}}$ for Coulomb interaction and $A(\omega) \propto
\omega^{(7+ (2.7 \nu)^{1/2} d/l)^{1/2}/4\nu-1}$ for short-range
interaction.  Complete 
numerical results are shown in Fig.~\ref{fig1}. They are in good
agreement with 
previous  calculations \cite {JohanssonKinaret94,Haussmann96,Halperin} as
well as with experimental data \cite{Eisenstein}.

{\it (ii) Incompressible case.} Within the
single-pole approximation
$K(q\to0) \simeq \Delta \rho_0/2 \alpha q^4$, where $\Delta$ is the
collective excitation gap at $q=0$, and $\alpha = (1 - \nu)/8 \nu$ 
is derived from the Laughlin
wavefunction at filling factors $\nu \equiv 2\pi l^2 \rho_0 =1/${\em odd
  integer}\cite{Girvin86}.  
 Using Eq.~(\ref {dispersion}) we find at long wavelength $\omega_q =
(\Delta \mu /2 \alpha \rho_0)^{1/2}$ independent of $q$. 
This formula can be used to deduce the value of $\mu$ if
$\Delta=\lim_{q\to0} \omega_{q}$ is known. Alternatively, one can
substitute $\mu=\mu_c$ from the classical Wigner crystal and obtain 
$\Delta_\nu=0.391  {\nu^{3/2}(1-\nu)^{-1}}{e^2 /l}$, which gives
$\Delta_{1/3}=0.11 e^2/l$, $\Delta_{1/5}=0.044 e^2/l$ and
$\Delta_{1/7}= 0.025 e^2/l$. 
These results are in good agreement
with variational estimates \cite{Girvin86,Jain}, with the
exception of the first one which is almost 30\%\ lower than the
variational one, but compares favorably with exact diagonalization
studies \cite{Yoshioka}.

The calculation of the spectral function is more subtle.
Straightforward application of the linear response formulas
(\ref{A}-\ref{Mtildehomogeneous}) 
is incorrect, because the addition (or removal) of
charge at point $\vec R$ creates a compressible region in the middle of
the liquid, changing the {\it topology} of the
incompressible region from simply to doubly connected.
The change in
topology can be taken into account in the following
manner.  We stipulate that the bulk
modulus kernel $-\rho_0^{2}\tilde \chi^{-1}(\vec r,\vec {r'})$ has the
form characteristic of an incompressible liquid (namely, $\vert \vec r -
\vec {r'}\vert^2 \log\vert \vec r - \vec {r'}\vert$) when both $\vec r$
and $\vec {r'}$ are within the incompressible region, but it is given by
the local form $K_c \delta(\vec r - \vec {r'})$ when either $\vec r$ or
$\vec {r'}$ are within the compressible ``core" of the excitation.   
Because the size of the core region is microscopic, its presence does
not affect significantly the frequencies of the long wavelength modes.
On the other hand, in the ``electron-phonon"
matrix element, given by Eq.~(\ref{Mtildehomogeneous}), we must use
$K_{\vec r}(q) \simeq K_c$, because $\vec r$ is inside the core region. 
Within this scheme, the calculation of the spectral
function can be straightforwardly carried out, without adjustable
parameters.  In Fig.~\ref{fig1}, we plot the results for $\nu=1/3$. 
We have used $q$-dependent $K$ and $\mu$ in order to fit accurately the
collective mode dispersion and structure factor \cite{Girvin86} at finite $q$. 
The essential difference between this and the compressible case is the
appearance of a $\delta$-function peak at $\omega=0$, which now
corresponds to $\mu_+$
(recall that in the incompressible liquid
 the chemical potentials for addition or removal of charge, $\mu_+$
and $\mu_-$, are different\cite{MorfHalperin}). The strength of the
$\delta$-function is 
$Z = \exp[ -\sum_q \vert M_q\vert^2/\omega_q^2]$, which would have vanished in
the compressible case,  and {\it does not vanish here because of the gap}.
The peak at $\mu_+$ reflects the ability of the
incompressible liquid to
accommodate the incoming electron in the ground state as topological
defects of the initial incompressible state, without creating
collective excitations.
The incoherent part of the spectral function, at higher frequencies,
corresponds to the creation of additional collective excitations
\cite{Footnote6}.

(2) {\it Edge dynamics}. The case of a smooth
compressible edge has  been treated in
previous publications \cite{CV96CV97drs}. Here we focus on the case of a
{\it sharp} edge, which is directly relevant to the interpretation of
recent {\it lateral} tunneling experiments by Chang {\em et al.} and
Grayson {\em et al.} \cite {Chang}. Let us consider a straight edge along
the $y$ axis, and let $eE \hat x$ be the gradient of the confinement
potential at the edge. The density $\rho_0$ is assumed to be uniform for
$x<0$, and  zero for $x>0$. The presence of the edge  electric field
breaks the rotational symmetry of the Hamiltonian
~(\ref{Hamiltonian}), and must be taken into account with the
additional term
\begin {equation}
H^{\mathrm{edge}} = e\int \rho (\vec r) \vec u(\vec r)\cdot 
\vec E\, d\vec r \,. 
\label {edge} \end {equation} 
With this term included into the 
eigenvalue problem (\ref{eigenvalueproblem}), we obtain a new set of
solutions, which satisfy the
conditions $\vec \nabla \cdot \vec g = 0$ and $\vec \nabla \times \vec g
= 0$ for $x < 0$. Neglecting the logarithmic correction arising
from the long-range Coulomb interaction, they obey the equation of motion
\begin {equation}
-i \omega g_\alpha(\vec r) = 
v \epsilon_{\alpha \beta} \partial_\beta g_x (\vec r)\,,
 \label{edgemotion} \end {equation}
where $v=cE/B$ is the classical drift velocity. 
The orthonormal solutions  are (for $x<0$) 
\begin {equation} \vec g_q(\vec r)
= q^{1/2} e^{qx-iqy}(\hat x - i \hat y)\,,
\label{edgewaves}\end{equation}
where $q>0$ is a one-dimensional  vector along the edge, and the eigenvalues
are $\omega_q = vq$. These are analogous to gravity waves on the
surface of a liquid.
 Because there is neither density change nor
shear strain  in the interior of the system, these solutions do not
depend on the values of the bulk elastic constants. 
The use of elasticity theory is justified at small $q$ since
the displacement field $\vec u$ is slowly varying. 

The effective edge dynamics can be derived from the full dynamics by
projecting the latter onto the subspace spanned by the edge-wave
solutions of Eq.~(\ref{edgewaves}). Within this subspace, we can define
an ``edge density" operator $\rho^{\mathrm{edge}} (y) \equiv \int  \delta \bar
\rho(x,y)dx = \rho_0 \bar u_x(0,y)$, where the bar denotes projection
onto the edge-wave subspace, i.e., for instance, $\bar u_x (0,y) \equiv
(l/ \rho_0^{1/2})\sum_{q>0} q^{1/2}[b_q e^{iqy} + b^\dagger_qe^{-iqy}]$.
It is easy to verify that the edge density satisfies the standard
Kac-Moody algebra \cite{Wen} $[\rho^{\mathrm{edge}}_q, 
\rho^{\mathrm{edge}}_{-q'}] = (\nu
q/2\pi)\delta_{qq'}$ where $\nu$ is the bulk filling
factor. Thus, we have  deduced the dynamics of the chiral
Luttinger liquid at the edge  from a projection of the canonical
dynamics of displacement fields in the bulk.  The
edge tunneling operator, obtained by imposing $[\psi(y), 
\rho^{\mathrm{edge}}(y')]
= \psi(y) \delta(y-y')$ in the edge-wave subspace, is still given by
Eq.~(\ref{psi}), with the sum running over the edge modes, and the matrix
element  $ M_q (y) = v(2 \pi q/\nu)^{1/2}e^{-iqy}$. Use of
Eqs.~(\ref{A}) and (\ref{g}) then leads to the conclusion that the
tunneling current must vanish, at low bias $V$, as $V^{1/\nu}$.   This
result is in good agreement with the experimental findings. The
present derivation explains why the tunneling exponent is found to
depend {\it only} on the bulk {\em density}, and not on whether the bulk is
compressible or not.  

In summary, we have developed a magneto-elastic bosonization scheme 
for the long wavelength  dynamics of the 2DEG in the LLL. Our results
show that this scheme can provide a unified description of different
physical effects in the bulk and at the edge of the system.

This work was supported by NSF grant No. DMR-9706788. One of
us (GV) wishes to thank I. Aleiner and L. Glazman for a useful
discussion at the beginning of this work.


\begin{references}
  
\bibitem{Bosonization}S. Tomonaga, Prog. Theor. Phys. {\bf 5}, 544 (1950);
J. M. Luttinger, J. Math. Phys. {\bf 4}, 1154 (1963); D. Mattis and
E. Lieb, J. Math. Phys. {\bf 6}, 304 (1965); A. Luther, Phys. Rev. B
{\bf 14}, 2153 (1975);  F. D. M. Haldane, J. Phys. C {\bf 14}, 2585
(1981). 

\bibitem{FradkinLuther} A. Luther, Phys. Rev. B {\bf 19}, 320
(1979); F. D. M. Haldane, Varenna Lectures (1992) and Helv. Phys. Acta
{\bf 65}, 152 (1992); A. H. Castro Neto and E. Fradkin, Phys. Rev. B
{\bf 49}, 10877 (1994).
   
\bibitem{Wen} X.~G. Wen, Phys. Rev. B {\bf 41}, 12838 (1990); {\em ibidem}
  {\bf 44}, 5708 (1991); Int. Journ. Mod. Phys. B {\bf 6}, 1711
  (1992).

\bibitem {Orgad} U. Z\"ulicke and A. H. MacDonald, Phys. Rev. B {\bf 54},
8349 (1996); Dror Orgad, Phys. Rev. Lett. {\bf 79}, 475 (1997).
 
\bibitem{Laughlin} R.~B. Laughlin, Phys. Rev. B {\bf 23}, 5632 (1981).

\bibitem{Girvin86} S.~M. Girvin, A.~H. MacDonald, and P.~M. Platzman,
  Phys. Rev. B {\bf 33}, 2481 (1986).
\bibitem {ZhangRead} S.-C. Zhang, H. Hansson and S. Kivelson, Phys.
Rev. Lett. {\bf 62}, 82 (1989); N. Read, {\it ibidem} {\bf 62}, 86
  (1989); D. H. Lee and M. P. A. Fisher, {\it ibidem} {\bf 63}, 903 (1989). 

\bibitem{JohanssonKinaret94} P. Johansson and J. M. Kinaret, Phys. Rev.
  B {\bf 50}, 4671 (1994).

\bibitem{Haussmann96} R. Haussmann, Phys. Rev. B {\bf 53}, 7357
  (1996).

\bibitem{Aleiner94} I.~L. Aleiner and L.~I. Glazman, Phys. Rev. Lett.
 {\bf 72}, 2935 (1994).

\bibitem{CV96CV97drs} S. Conti and G. Vignale, Phys. Rev. B {\bf 54},
  14309 (1996); and Phys. E {\bf 1}, 101 (1998),
  cond-mat/9709055. 

\bibitem{HanThouless} J.~H. Han and D.~J. Thouless, Phys. Rev. B {\bf
    55}, 1926 (1997); J.~H. Han, {\em ibidem} {\bf 56}, 15806 (1997). 

\bibitem{Aleiner95} I.~L. Aleiner, H.~U. Baranger,
and L.~I. Glazman, Phys. Rev. Lett. {\bf 74}, 3435 (1995);
H. Westfahl, A.~H. Castro Neto, and A.~O.
Caldeira, Phys. Rev. B {\bf 55}, 7347 (1997).

\bibitem{Chang}
A.~M. Chang, L.~N. Pfeiffer, and K.~W. West, Phys. Rev. Lett. {\bf
77},  2538 (1996) and Grayson {\em et al.} to be published.

\bibitem{Nozieres}
P. Nozi\`eres, {\em The theory of interacting Fermi
systems} (W.~A. Benjamin, New York, 1964).

\bibitem{Landau7} L.~D. Landau and E. Lifshitz, {\em Theory of
 Elasticity}, Vol.~7 of {\em Course of theoretical Physics}, 3rd
 ed. (Pergamon Press, Oxford, 1986).

\bibitem{Footnote11} The extra sign is caused by the property of the
  scalar product $(f,f)=-(f^*,f^*)$ together with the definition
  $g_{-n}=g_n^*$. The fact that the positive frequency eigenfunctions
  ($n>0$) are the ones with positive norm follows from the stability
  of the ground-state.

\bibitem{Mahan}
G.~D. Mahan, {\em Many-particle Physics} (Plenum Press, New York, 1990).

\bibitem {Maradudin} L. Bonsall and A. A. Maradudin, Phys. Rev. B {\bf
    15}, 1959 (1977).

\bibitem {Halperin} Y. Hatsugai, P.-A. Bares, and X. G. Wen, Phys.
Rev. Lett. {\bf 71}, 424 (1993); Song He, P. M. Platzman, and B. I.
Halperin, {\it ibidem} {\bf 71}, 777 (1993).

\bibitem{Eisenstein} J.~P. Eisenstein, L.~N. Pfeiffer, K.~W. West,
Phys. Rev. Lett. {\bf 69}, 3804 (1992).

\bibitem {Jain} J.~K. Jain and R.~K. Kamilla, Int. J. Mod. Phys. {\bf 11},
  2621 (1997) and references therein.

\bibitem{Yoshioka} D. Yoshioka, J. Phys. Soc. Jap. {\bf 55}, 885 (1986).

\bibitem{MorfHalperin} R. Morf and B.~I. Halperin, Phys. Rev. B {\bf
    33}, 2221 (1986).
 
\bibitem {Footnote6} Neither the peak at $\omega = \mu_+$ nor the oscillatory
behavior at higher frequency were seen in the experiment of Ref.
\cite{Eisenstein}. A possible explanation is that tunneling took place
primarily between compressible regions.  This would be consistent with the
observed linear dependence of the density on bias voltage.

\end{references}
\end {document}